\documentclass[RNAAS]{aastex63}

\newcommand{\approxgt}{\mbox{$\;^{>}\hspace{-0.24cm}_{\sim}\;$}}

\graphicspath{{./}{figures/}}
\begin{document}

\title{Reprocessing of a Green Bank 43-meter Telescope Survey of Unidentified Bright Radio Sources for Pulsars and Radio Bursts}

\correspondingauthor{Fronefield Crawford}
\email{fcrawfor@fandm.edu}

\author[0000-0002-2578-0360]{Fronefield Crawford}
\affiliation{Department of Physics and Astronomy, Franklin and Marshall College, P.O. Box 3003, Lancaster, PA 17604, USA}

\author{James Margeson}
\affiliation{Department of Physics and Astronomy, Franklin and Marshall College, P.O. Box 3003, Lancaster, PA 17604, USA}

\author{Benjamin Nguyen}
\affiliation{Department of Physics and Astronomy, Franklin and Marshall College, P.O. Box 3003, Lancaster, PA 17604, USA}

\author{Tanya Saigal} 
\affiliation{Department of Physics and Astronomy, Franklin and Marshall College, P.O. Box 3003, Lancaster, PA 17604, USA}

\author{Olivia Young}
\affiliation{Department of Physics and Astronomy, West Virginia University, P.O. Box 6315, Morgantown, WV 26506, USA}
\affiliation{Center for Gravitational Waves and Cosmology, West Virginia University, Chestnut Ridge Research Building, Morgantown, WV 26505, USA}
\affiliation{School of Physics and Astronomy, Rochester Institute of Technology, Rochester, NY 14623, USA}

\author[0000-0003-0385-491X]{Devansh Agarwal}
\affiliation{Department of Physics and Astronomy, West Virginia University, P.O. Box 6315, Morgantown, WV 26506, USA}
\affiliation{Center for Gravitational Waves and Cosmology, West Virginia University, Chestnut Ridge Research Building, Morgantown, WV 26505, USA}

\author[0000-0002-2059-0525]{Kshitij Aggarwal}
\affiliation{Department of Physics and Astronomy, West Virginia University, P.O. Box 6315, Morgantown, WV 26506, USA}
\affiliation{Center for Gravitational Waves and Cosmology, West Virginia University, Chestnut Ridge Research Building, Morgantown, WV 26505, USA}

\begin{abstract}
We have reprocessed a set of observations of 75 bright, unidentified, steep-spectrum polarized radio sources taken with the Green Bank 43-m telescope to find previously undetected sub-millisecond pulsars and radio bursts. The (null) results of the first search of these data were reported by \citet{scl+13}. Our reprocessing searched for single pulses out to a dispersion measure (DM) of 1000 pc cm$^{-3}$ which were classified using the Deep Learning based classifier \textsc{fetch}. We also searched for periodicities at a wider range of DMs and accelerations. Our search was sensitive to highly accelerated, rapidly rotating pulsars (including sub-millisecond pulsars) in compact binary systems as well as to highly-dispersed impulsive signals, such as fast radio bursts. No pulsars or astrophysical burst signals were found in the reprocessing.
\end{abstract}


\section{Introduction} 

\citet{scl+13} observed a set of 75 bright, unidentified, and unresolved steep-spectrum radio sources with the Green Bank 43-m telescope that were identified in the NRAO VLA Sky Survey \citep{ccg+98} as having significant linear polarization (see Table 1 of \citet{scl+13} for a list of the observed targets). These observations were undertaken to search for sub-millisecond pulsars, the existence of which has been predicted by theoretical models and the detection of which would provide important constraints on the neutron star equation of state \citep{lp01}. Such rapidly spinning pulsars could have been missed in previous large-scale sky surveys for pulsars owing to the limited sampling rates and bandwidth channelization of such surveys. The 75 target sources were each observed with the 43-m telescope for 900 s at a center frequency of 1200 MHz with the WUPPI 800 MHz wide-bandwidth backend \citep{mml+12}. The observations used very fast sampling (61.44 $\mu s$) and narrow frequency channels (4096 channels of width 195 kHz) which preserved sensitivity to pulsars down to sub-millisecond periods (see Fig.~1 of \citealt{scl+13}). Note that these same sources were also previously observed with the Lovell Telescope at 610 MHz with very fast sampling, but with only a very small total bandwidth (1 MHz). No discoveries were made in that search \citep{ckb00}. The 43-m telescope data were originally processed by \citet{scl+13} using the \textsc{presto}\footnote{https://www.cv.nrao.edu/$\sim$sransom/presto/} \citep{r01,rem02} software package to search for periodicities and the \textsc{sigproc}\footnote{http://sigproc.sourceforge.net/} \citep{l11} package to search for single pulses. In both processing runs, the data were searched for signals at dispersion measures (DMs) ranging from 0 to 100 pc cm$^{-3}$. Since the targets were mostly located far from the Galactic plane, this DM range encompassed the maximum Galactic DM that would be expected for almost all of the target lines of sight, according to the NE2001 Galactic electron model \citep{cl02}. \citet{scl+13} searched for periodic signals at moderate accelerations, up to $\pm 2.3~(P/1~{\rm ms})$ m s$^{-2}$ (where $P$ is the pulsar spin period), to account for binary motion. Their sensitivity threshold to pulsations was $\sim 3$ mJy for $P \approxgt 1$ ms (see Fig.~1 of \citealt{scl+13}). This is significantly below the flux densities of the target sources (estimated to be $> 20$ mJy at 1200 MHz in all cases), indicating that sensitivity was not a limiting factor in their non-detections. No new astrophysical signals were confirmed in their search.

\section{Analysis} 

We have undertaken an expanded reprocessing of this data set where we have searched for signals at a wider DM and acceleration range in several passes through the data. This was motivated by two considerations. The first is that highly dispersed bursts such as extragalactic fast radio bursts \citep[FRBs; e.g.,][]{phl19,cc19} or highly dispersed Galactic sources (either periodic or transient) could be present in the data (although these latter sources are unlikely to be present given the relatively small expected contributions from Galactic electrons to the DM along the target lines of sight). Such highly dispersed signals would have been missed in the original processing. The second is that very highly accelerated, rapidly rotating binary pulsars may have been missed if their accelerations significantly exceeded the range searched in the original processing.

In our reprocessing of the data, we searched for periodic sources using \textsc{sigproc} in three stages that each encompassed different DM and acceleration ranges. We also conducted a new search for single pulses using \textsc{heimdall}\footnote{https://sourceforge.net/projects/heimdall-astro/} \citep{b12}
and classified the single-pulse candidates using the convolutional neural network based classifier \textsc{fetch}\footnote{https://github.com/devanshkv/fetch} \citep{aab+20}. Table~\ref{tbl-1} outlines these different processing passes. In the first pass, we searched for periodicities out to a DM of 1000 pc cm$^{-3}$ but with no acceleration search. We combined pairs of time samples and frequency channels in the raw data before processing, thereby reducing the time and frequency resolution by a factor of two. In the second pass, we searched an acceleration range $\pm 12$ m s$^{-2}$ out to a DM of 200 pc cm$^{-3}$ using this same reduced time and frequency resolution. For the third pass, we searched for very highly accelerated systems (out to $\pm 1000$ m s$^{-2}$) with the native sampling and channelization, but with no DM search (i.e., we searched only at a trial DM of 0). In this third pass, pulsars would only be detectable if they were very nearby with very small DMs. Given the brightness of these sources (possibly indicating small distances) and given that the expected DM for our target sources would be DM $< 5$ pc cm$^{-3}$ for distances less than 0.2 kpc \citep{cl02}, it is reasonable to focus on small DMs for this particular processing pass (although large scintillation bandwidths at such small distances cannot be completely discounted as a factor for some non-detections). Such highly accelerated pulsars would be present in extremely compact binary systems having orbital periods of order an hour or less. These systems are favored as possible hosts of sub-millisecond pulsars \citep{bpd+01}, which are exactly the kinds of pulsars we were hoping to discover. For the single-pulse search, we searched the full-resolution data out to a DM of 1000 pc cm$^{-3}$, a range in which FRBs might be present. In this search,  \textsc{heimdall} applied boxcar matched filters with widths up to 512 samples to each dedispersed time series in order to maximize sensitivity to pulses in our data having widths up to 30 ms. \textsc{heimdall} produced a set of candidates that were then passed to \textsc{fetch} to determine the likelihood of each detected candidate signal being a real, astrophysical pulse. As a check, a large number of individual giant pulses from a diagnostic observation of the Crab pulsar taken with the 43-m telescope were easily detected by \textsc{heimdall} and categorized as being real by \textsc{fetch}. 

\section{Results}

We found no convincing pulsar candidates or astrophysical burst signals in any of the four reprocessing passes through the data. Our sensitivity limit to periodic sources remains $\sim 3$~mJy for most pulsar spin periods for relatively small DMs \citep{scl+13}. Our peak-flux detection limit for single pulses is $\sim 8$~Jy for a 1 ms pulse width, assuming a signal-to-noise detection threshold of 7 (the threshold we used in our search). The upper limit for the all-sky FRB event rate (derived from the absence of any FRBs detected in the 18.75 hr of on-sky time in the DM range we searched) is several orders of magnitude higher than rate limits reported from other, more extensive surveys \citep[e.g.,][]{rlb+16,crt+16} and is therefore not constraining.

\begin{deluxetable}{ccccc}
\tablecaption{Search Parameters Used in Reprocessing\label{tbl-1}}
\tablehead{
\colhead{Software Package} \vspace{-0.2cm} & \colhead{Sampling} & \colhead{Number of} & \colhead{Maximum} &  \colhead{Acceleration} 
\\
\colhead{(Search Type)} & \colhead{Time ($\mu s$)} & \colhead{Frequency Channels} & \colhead{DM (pc cm$^{-3}$)} &  \colhead{Range (m s$^{-2}$)}
}
\startdata
\textsc{sigproc} (Periodicity)          & 122.88 & 2048 & 1000 & 0  \\
\textsc{sigproc} (Periodicity)          & 122.88 & 2048 & 200  & $\pm 12$  \\
\textsc{sigproc} (Periodicity)          & 61.44  & 4096 & 0    & $\pm 1000$  \\
\textsc{heimdall}/\textsc{fetch} (Single Pulse)  & 61.44  & 4096 & 1000 & N/A 
\enddata
\tablecomments{All observations had a center frequency of 1200 MHz, a bandwidth of 800 MHz, and an integration time of 900 s. Further observational details are presented in \citet{scl+13}.}
\end{deluxetable}  

%

\end{document}